# Realistic atomistic structure of amorphous silicon from machine-learning-driven molecular dynamics

Volker L. Deringer,[a,b,*] Noam Bernstein,[c] Albert P. Bartók,[d] Matthew J. Cliffe,[b] Rachel N. Kerber,[b] Lauren E. Marbella,[b] Clare P. Grey,[b] Stephen R. Elliott,[b] and Gábor Csányi[a]

[a]Department of Engineering, University of Cambridge, Cambridge CB2 1PZ, United Kingdom
[b]Department of Chemistry, University of Cambridge, Cambridge CB2 1EW, United Kingdom
[c]Center for Materials Physics and Technology, U.S. Naval Research Laboratory, Washington, DC 20375, United States
[d]Scientific Computing Department, Science and Technology Facilities Council, Rutherford Appleton Laboratory, Oxfordshire OX11 0QX, United Kingdom

*E-mail: vld24@cam.ac.uk

**Amorphous silicon (*a*-Si) is a widely studied non-crystalline material, and yet the subtle details of its atomistic structure are still unclear. Here, we show that accurate structural models of *a*-Si can be obtained by harnessing the power of machine-learning algorithms to create interatomic potentials. Our best *a*-Si network is obtained by cooling from the melt in molecular-dynamics simulations, at a rate of $10^{11}$ K/s (that is, on the 10 ns timescale). This structure shows a defect concentration of below 2% and agrees with experiments regarding excess energies, diffraction data, as well as $^{29}$Si solid-state NMR chemical shifts. We show that this level of quality is impossible to achieve with faster quench simulations. We then generate a 4,096-atom system which correctly reproduces the magnitude of the first sharp diffraction peak (FSDP) in the structure factor, achieving the closest agreement with experiments to date. Our study demonstrates the broader impact of machine-learning interatomic potentials for elucidating accurate structures and properties of amorphous functional materials.**

Amorphous silicon (*a*-Si) is among the most fundamental and widely studied of non-crystalline materials, with applications ranging from photovoltaics and thin-film transistors to electrodes in batteries (1–5). Its atomic-scale structure is traditionally approximated in a Zachariasen-like picture (6) with all atoms in locally "crystal-like", tetrahedral environments, but without long-range order (7–9). However, the real material contains a non-zero amount of coordination defects, colloquially referred to as "dangling bonds" (under-coordinated sites) and "floating bonds" (over-coordinated sites). Knowing the properties and abundance of such defects is important, as they can control electronic and other macroscopic properties. We note at the outset that, although defect sites in *a*-Si may be passivated by hydrogenation (to give "*a*-Si:H") in some synthetic conditions, we here focus on the archetypical, hydrogen-free material as made in ion-implantation or sputter-deposition experiments (10–14).

Even the most advanced experimental approaches do not directly allow the observation of the bulk atomic structure in amorphous materials. Despite significant advances, including *in situ* NMR techniques (15, 16) and "inverse" approaches such as Reverse Monte Carlo (RMC) modeling of diffraction data (17, 18), only indirect knowledge can be gained about the local atomic environments, and that only in a statistical sense. For almost three decades, molecular-dynamics (MD) simulations have therefore played a crucial and complementary role in the field, with *a*-Si being a prominent example (19–23). These simulations either use density-functional theory (DFT) or classical force fields, which both have clear advantages but also inherent drawbacks. DFT-MD describes a system with quantum-mechanical accuracy and can largely correctly capture the structural and bonding subtleties of liquid and amorphous materials. However, it is highly computationally expensive, and therefore it allows only limited system sizes (a few hundred atoms at most) and timescales to be simulated. Indeed, the cooling rates that have previously been reported in DFT-MD simulations of *a*-Si ($\approx 10^{14}$ K/s) are several orders of magnitude faster than those in experiments (19–21). On the other hand, classical force fields require much less computational effort, giving access to nanometer-scale ("device-size") systems (9), but they are rarely accurate enough to fully correctly describe the structural variations present in the amorphous state.

Capitalizing on today's "big-data" revolution, an emerging line of research is the use of machine-learning (ML) algorithms to speed up atomistic simulations (24–28). By "learning from" (or, more accurately, *fitting to*) quantum-mechanical reference data for energies and forces, ML-based interatomic potentials can enable simulations with an accuracy that is largely comparable to DFT, but with a computational cost that is orders of magnitude lower, and with linear (order-*N*) scaling behavior. We believe that such ML potentials are particularly promising for disordered and amorphous functional materials, which must be represented by nanometer-scale structural models containing several hundreds or thousands of atoms. A landmark example has been the development of an artificial neural-network potential for the phase-change material GeTe (29). The structural complexity in its amorphous phase is a challenge for simulations and requires DFT-level accuracy (30). The unique combination of speed and accuracy afforded by ML potentials has now enabled simulation of the crystallization properties of GeTe (31) including entire nanowires (32). We recently introduced an ML potential for amorphous carbon (33), based on the Gaussian approximation potential (GAP) framework (25) and the Smooth Overlap of Atomic Positions (SOAP) atomic similarity kernel (34), which captures the intricate structural, mechanical, and surface properties of the material. Very recent work using artificial neural-network potentials has allowed for the atomistic modeling of amorphous Li$_x$Si phases relevant in battery applications (35). Finally, such potentials have been used in seminal studies to describe the complex phase transitions between polymorphs of *crystalline* Si (24, 36).

In this work, we show how realistic atomistic modeling of *a*-Si can be enabled by an ML-based interatomic potential, again using SOAP and GAP. We first report on melt–quench simulations with cooling rates much slower (that is, better) than what can be achieved in quantum-mechanical-based simulations, and we show how this leads to a higher-quality and lower-energy structure of *a*-Si. Our structural models show excellent agreement with experiments, including $^{29}$Si NMR shifts and diffraction data for high-quality samples, and open the door for future combined modeling and experimental studies on disordered and amorphous materials.

## Results and discussion

Simulated quenching from the melt has become a widely used technique for generating amorphous model networks. In such MD simulations, one starts with a liquid and progressively lowers the temperature, "freezing in" the atomic configurations into an amorphous structure. However, for silicon, this approach is not trivial, due to the change in coordination environments between the high-coordination metallic liquid and the tetrahedral-like amorphous state. We decided to perform a set of variable-volume and constant-pressure (NPT) quench simulations in which we varied the quench rate, and thus the run-time, by several orders of magnitude. For the moment, we focus on a system size of 512 atoms in the simulation cell. This is significantly larger than what has so far been accessible to DFT (64–216 atoms) (19–21, 23), but smaller than what is possible for empirical potentials; this question will be addressed directly later on. Our fastest quench rate ($10^{14}$ K/s) corresponds to early, seminal DFT studies (19–21), whereas our simulations at $10^{12}$ K/s mirror the limit of what is presently possible for *ab initio* quality MD. In contrast, using our recently developed GAP model (28) here allows us to increase the simulation time tenfold beyond that, namely decreasing the quench rate to $10^{11}$ K/s, while retaining similar accuracy.

While an increase in simulation time by one order of magnitude may seem incremental at first sight, the full power of ML potentials becomes apparent when looking at the overall computational effort required (Fig. 1A). For demonstration, we performed a brief DFT- MD simulation on a 512-atom *a*-Si network and use the timing information for a rough estimate of the cost for a full simulation (Fig. 1A), as detailed in the Supporting Information. Extrapolating from our data, a quench with a rate of $10^{11}$ K/s at this system size will require around 16 million core hours, which corresponds to occupying 576 processor cores on a state-of-the-art supercomputer for over 3 years. Using the current nominal costs for the UK national supercomputing service, ARCHER, this is equivalent to costs of $185,000. In contrast, the same quench rate in our GAP-MD simulations required below 40,000 core hours, equivalent to nominal costs below $500. Using GAP, it would hence be possible to decrease the quench rate even further, but given the results obtained at $10^{11}$ K/s we subsequently chose to increase the system size instead (see below).

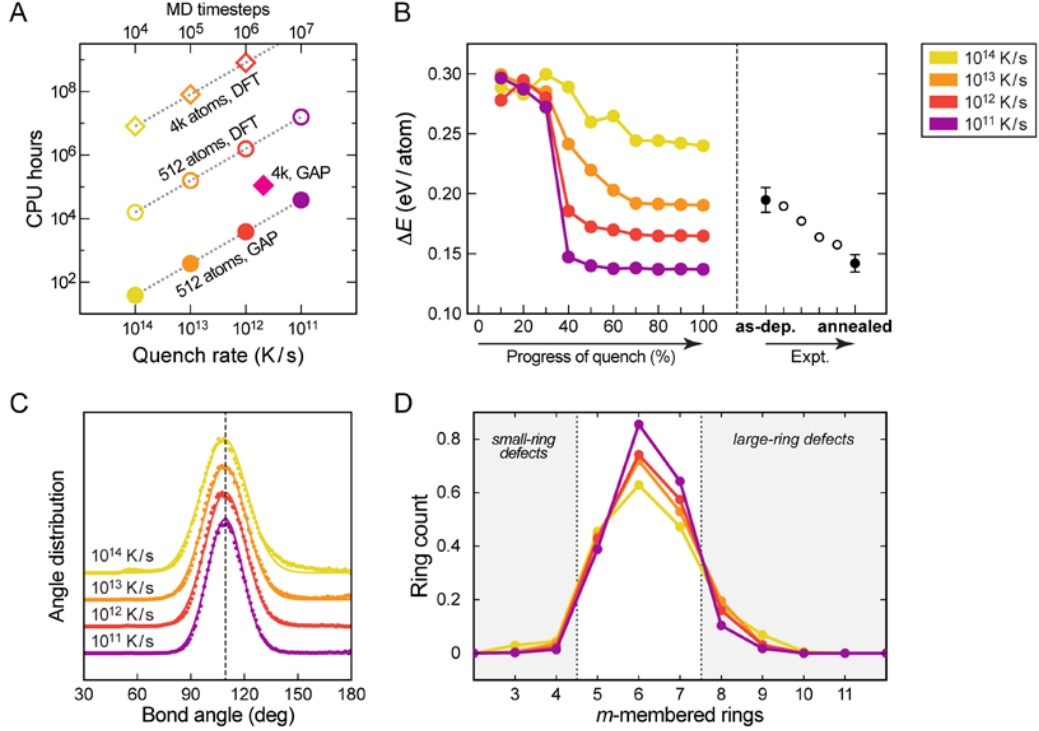

**Fig. 1.** Unlocking slow quenching in molecular-dynamics simulations, using the GAP machine-learning framework, and its application to *a*-Si. (A) Computing time required for DFT- and GAP-MD simulations with different quench rates. Decreasing the quench rate by an order of magnitude increases the number of required MD steps, and thus the CPU cost, by the same factor. Quench rates of $\approx 10^{12}$ K/s ($\approx$ one million steps) have so far been the limit for 512-atom DFT-MD simulations, and a system size of 4,096 atoms ("4k") has been widely out of reach. Both limits can be overcome using GAP, due to its significantly lower computational cost and its linear scaling with system size. DFT timing information has been extrapolated from a short trajectory (Supporting Information). (B) Stability of 512-atom *a*-Si structures, taken at various stages of GAP melt–quench trajectories and subsequently relaxed; energies are given relative to crystalline (diamond-type) Si. Experimental data refer to samples that have been freshly deposited ("as-dep.") or annealed at progressively higher temperatures (12). (C) Angle distribution functions for *a*-Si GAP structures obtained at different quench rates. Points show original data, sampled from short (5 ps) MD simulations; lines show Gaussian fits. (D) Medium-range order in these *a*-Si networks, assessed by shortest-path ring statistics (37). We label rings containing fewer than five atoms as "small-ring defects", and rings containing more than seven atoms as "large-ring defects".

The slow quench rate of $10^{11}$ K/s, "unlocked" here using GAP, is indeed required to generate reliable structural models of *a*-Si. This is seen in Fig. 1B: we took structural snapshots at various increments of the quench simulations, optimized them into local minima, and plotted their energy (relative to the thermodynamically stable form, diamond-type *c*-Si) as a function of how far the quench has progressed in time from the liquid to the amorphous state. Data points at "100%" therefore correspond to the final *a*-Si structures. The right-hand side shows the experimental sample stability with increasing annealing and thus ordering, based on calorimetry (as is common, we approximate $\Delta E \cong \Delta H$ when comparing theory and experiment). Quenches at intermediate rates lead to *a*-Si networks that are as stable as freshly deposited or partially annealed samples ($\Delta E \approx$ 0.17–0.20 eV/atom). In contrast, our slowest quench at $10^{11}$ K/s yields a structure whose stability matches the experimental result for a well-annealed sample from Ref. (12) ($\Delta E \approx 0.14$ eV/atom).

The benefit of slow quenching is further seen in two of the most common structural indicators used for amorphous solids. In *a*-Si, the bond angles are distributed around the ideal tetrahedral value (109.5°; Fig. 1C). Fitting Gaussian distributions to these data allows us to determine the full width at half maximum (FWHM), which decreases gradually from 30° to 22° with increasingly slower quenching. The experimental value for the bond-angle deviation of ≈ 11° (Ref. (10)), if taken to correspond to the half width at half maximum of a Gaussian distribution, is consistent with our computed value for the slowest quench. Moreover, the medium-range structural order is important in covalent amorphous networks (38), and we quantify it here using shortest-path ring statistics (Fig. 1D) (37). In diamond-type *c*-Si, all atoms are in six-membered ring (cyclohexane-like) configurations, whereas *a*-Si also contains a large number of five- and seven-membered rings, as well as a lesser amount of smaller and larger rings. Again, progressively increasing structural order is seen with slower quenching rates.

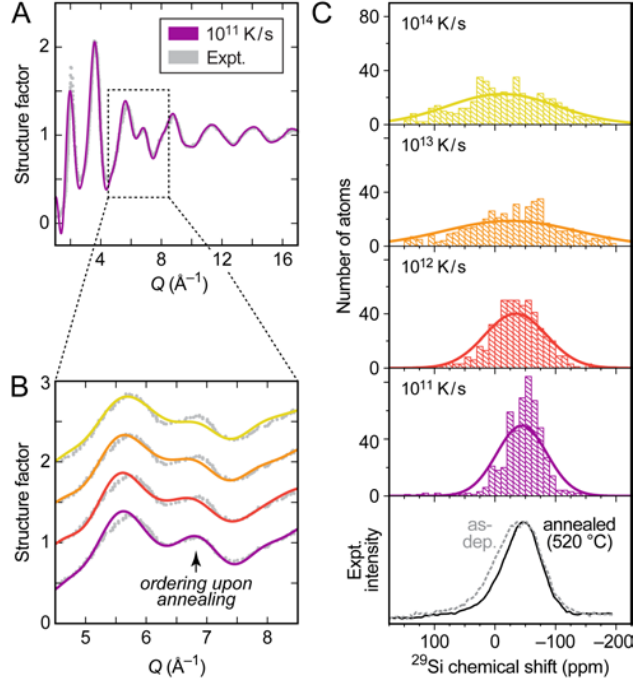

**Fig. 2.** Progressive ordering in *a*-Si, comparing 512-atom GAP structures obtained with various quench rates to experimental data. (A) Computed structure factor $S(Q)$ (purple) compared to X-ray diffraction measurements for a well-annealed sample (gray; taken from Ref. (14)). (B) Close-up around the third peak, in which data for the different melt–quench simulations have been offset vertically (color code as in Fig. 1), and are each compared to the same experimental dataset (points) (14). (C) Solid-state $^{29}$Si NMR chemical shifts, computed for the GAP-quenched structures (raw data as histograms; Gaussian fits as colored lines), comparing to experimental data taken from Ref. (39) (bottom). For numerical values, see Table 1.

In Fig. 2A–B, we show computed structure factors, $S(Q)$, which can be compared to diffraction experiments. The third peak (at ≈ 5–7 Å$^{-1}$; enlarged in Fig. 2B) gradually splits into two well-defined sub-peaks when moving from the $10^{14}$ K/s (yellow) to the $10^{11}$ K/s quench-rate data (purple). This progressive peak splitting is consistent with experimental observations: as-deposited samples show a fairly featureless third peak, whereas annealed ones (and also our $10^{11}$ K/s result) exhibit a clear splitting into two sub-peaks (14). Even better agreement with the experimental structure factor can be achieved for a larger structural model containing 4,096 atoms, which we will show below.

We furthermore computed solid-state $^{29}$Si NMR chemical shifts, δ, for all atoms in the unit cells, thereby characterizing each local atomic environment individually. We use established DFT-based algorithms (40–42), as detailed in the Methods section, and reference all δ values to tetramethylsilane (TMS), analogous to experiments. The results for the different GAP-quenched structures are shown in Fig. 2C (histograms). Furthermore, due to the broad distribution of δ values in the amorphous state, we fit Gaussian profile functions to these data (lines). We compare the output of these computations to experiments for pure *a*-Si as prepared by sputter deposition (39). The latter samples were analyzed via secondary-ion mass spectrometry (SIMS), showing no measurable oxygen contamination and ≈ 0.2 at.-% hydrogen in the samples (39). This low level of impurity is thought to have little or no impact on the $^{29}$Si NMR results, enabling direct comparison to our simulations. In addition to the numerical values reported in Ref. (39), we fit a Gaussian profile to the experimental data for the sample annealed at 520 °C (before the onset of crystallization at higher temperature, which leads to the formation of a shoulder in the NMR signal). We perform this fit using the same numerical procedure as for our DFT data (Table 1). This allows us to quantify both the center of mass for the chemical shifts and the FWHM, yielding simple and well-defined numerical quality indicators that can be used to assess the quality of any given structural model.

Clearly, simulations using the two fastest quench rates (yellow and orange in Fig. 2C) lead to structures with very large scatter in the computed NMR shifts, as a direct consequence of their distorted atomic environments. The *a*-Si network generated at a slower quench rate, $10^{12}$ K/s (red), agrees more appreciably with experiment, with a center of mass for the Gaussian fit of $δ_{DFT}$ = –34 ppm, compared to $δ_{exp}$ = –38.3 ppm for as-deposited *a*-Si, and $δ_{exp}$ = –42.9 ppm for a sample that has been annealed at 580 ºC (39). In other words, there is a progressive shift to lower frequency in the experimental data with increasing structural ordering, and this is reproduced by our quenched structure at $10^{11}$ K/s ($δ_{DFT}$ = –44 ppm), both qualitatively and quantitatively (to within a few ppm).

A Gaussian fit to the experimental data (Table 1) yields very similar results, and it also shows that the signal width (FWHM) is too large except for the $10^{11}$ K/s quench.

**Table 1. Figures of merit for $^{29}$Si NMR parameters from simulation and experiment**

|  | δ (ppm) | Gaussian FWHM (ppm) |
|---|---|---|
| GAP-MD, $10^{14}$ K/s quench | –12 | 213 |
| GAP-MD, $10^{13}$ K/s quench | –25 | 257 |
| GAP-MD, $10^{12}$ K/s quench | –34 | 120 |
| GAP-MD, $10^{11}$ K/s quench | –44 | 97 |
| Expt. (as-dep.; Ref. (39)) | –38.3 | |
| Expt. (580 °C annealed; Ref. (39)) | –42.9 | |
| Expt. (520 °C annealed; Gaussian fit) | –45 | 76 |
| Relaxed WWW model | –54 | 68 |

For comparison, we performed a similar analysis for a current state-of-the-art structural model, namely, a DFT-optimized Wooten–Winer–Weaire (WWW) network of *a*-Si (43). This yields a narrower distribution (because the local structural fragments are more similar to one another; see below), but the computed chemical shift for this structure, $\delta_{DFT} = -54$ ppm, deviates visibly from the experimental data (Table 1).

We now place our melt–quench simulations into a wider context, as there are several different ways of modeling *a*-Si. First, we survey results of RMC modeling, which is an established means of extracting structural information from diffraction data (17). Recent work by some of us showed that reasonable restraints during refinement can improve the RMC description of *a*-Si (18). In particular, the SOAP similarity measure, initially developed to encode atomic structure in ML potentials (34), turned out to be a useful restraint for RMC refinements (43). Up to this stage, the SOAP-RMC output, subsequently relaxed using DFT, has been known to provide a high-quality structural model of *a*-Si (43). We now take the same structures but anneal them further using GAP: heating to 1,100 K, holding, and cooling back to 300 K, for a total simulation time of 50 ps. This relatively short annealing time is thought to be appropriate, as a recent DFT-MD study showed that annealing a quenched structure at 10 ps versus 20 ps had no appreciable effect on the outcome (23). We also performed the same annealing procedure for the DFT-optimized WWW model from Ref. (43); a somewhat similar strategy has been followed before, based on a tight-binding model and a system size of 216 atoms (22). Finally, we include a state-of-the-art 216-atom structure that was carefully generated in a recent work, by slow quenching using the empirical Tersoff potential and subsequent multistep optimization using DFT (here labeled "Tsf+DFT") (23).

We compare these structural models in Fig. 3 using three types of quality indicators. First, we report the number of coordination defects (Fig. 3A), counting three- and fivefold bonded atoms with a bond-length cutoff of 2.85 Å. We then measure the distortion from ideal tetrahedral coordination environments in two ways: by fitting Gaussians to the angle distributions as before and determining their FWHM, and by using a numerical order parameter (44) that was employed earlier for tetrahedral environments in liquid water (45) and chalcogenide glasses (46) (Fig. 3B). Beyond nearest-neighbor environments, we quantify the medium-range order using shortest-path ring statistics, as above (37), again considering as "defects" any rings with fewer than 5 or more than 7 members (Fig. 3C). In all cases, the GAP-quenched structure with the slowest quench rate ($10^{11}$ K/s) exhibits very good figures of merit.

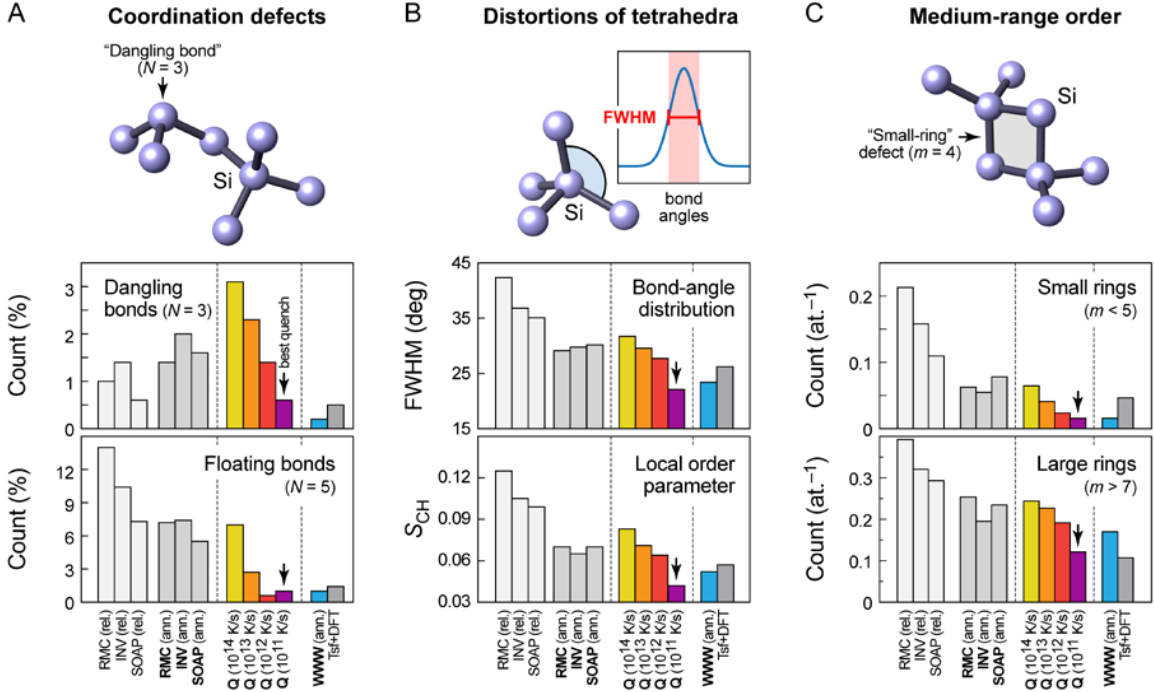

**Fig. 3.** A survey of *a*-Si structural models, using common quality criteria; for all of them, the lower, the better. (A) Coordination defects, reporting the count of "dangling bonds" (threefold bonded atoms) and "floating bonds" (fivefold bonded atoms). (B) Deviations from tetrahedral geometry, quantified via the FWHM of the fitted angle distribution (*top*; see also Fig. 1C) and a variant of the Chau–Hardwick order parameter, $S_{CH}$ (*bottom*) (44). (C) Count of small- and large-ring defects (see also Fig. 1D). Light gray bars refer to structures from Ref. (43), generated using pure reverse-Monte Carlo ("RMC"), INVERT restraints ("INV") (18), or SOAP restraints, respectively. Structures denoted "rel." have been DFT-relaxed in Ref. (43); structures denoted "ann." have been further annealed using GAP. "Q" denotes our GAP quenches at the different rates.

Finally, we prepared a larger *a*-Si structural model containing 4,096 atoms (Fig. 4A), using GAP-MD and a variable quench rate between $10^{11}$ and $10^{13}$ K/s, as detailed in the Methods section. This system size is comfortably in reach for ML-based interatomic potentials (31), as they are linearly scaling with system size due to their finite cutoff radius (cf. Fig. 1A). Having access to *ab initio* quality structural models on the 4-nm length scale allows us to study the medium-range order more closely. This fundamental question has been discussed in recent work on nearly hyper-uniform networks (47, 48), in particular, by quantifying the inverse height ($H^{-1}$) of the first sharp diffraction peak in the structure factor (or its computed analogue) at around 2 Å$^{-1}$. This quantity is taken as a measure for the degree of structural ordering (47, 48).

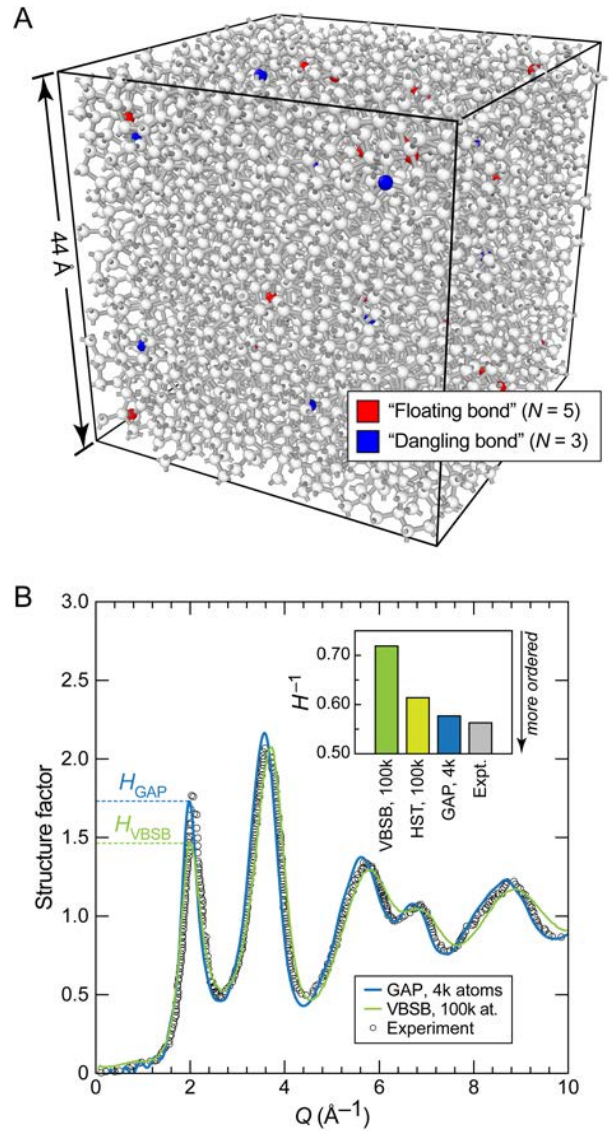

**Fig. 4.** A nanoscale structural model of *a*-Si containing 4,096 atoms, quenched using GAP-MD at a variable rate between $10^{11}$ and $10^{13}$ K/s (see Methods). (A) Ball-and-stick drawing, emphasizing the low concentration of coordination defects. The vast majority of atoms are fourfold coordinated ($N = 4$) and are shown as colorless spheres. (B) Reciprocal-space fingerprints in the structure factor, comparing to results for two of the largest structural models to date (containing 100,000 atoms, "100k") and to experimental data from Ref. (14). The height of the first sharp diffraction peak, $H$, serves as an indicator for structural ordering; in accord with previous literature, we plot $H^{-1}$ values in the inset. The structure labeled "VBSB" is taken from Vink et al. (9); the structure labeled "HST" is from a more recent study by Hejna, Steinhardt, and Torquato (47).

We compare our structure with the current state of the art, viz. *a*-Si systems containing 100,000 atoms (9, 49), in Fig. 4B. Surprisingly, the latter ultra-large system size does *not* seem to be needed if the structural modeling itself is sufficiently accurate. Indeed, looking at $H^{-1}$, our GAP approach outperforms the previous simulation results in much larger cells, and leads, again, to almost quantitative agreement with experiment ($H^{-1}$ = 0.58 with GAP, $H^{-1}$ = 0.57 in experiment; Fig. 4B). By comparison, an *a*-Si structural model of the same size (4,096 atoms) but generated using empirical potentials gave a much larger $H^{-1}$ = 0.81 (Ref. (49)). Moreover, our slowest-quenched GAP-based system, even smaller with 512 atoms/cell, yields $H^{-1}$ = 0.66, remarkably still outperforming the 100,000-atom structure from Ref. (14) ($H^{-1}$ = 0.68). Looking beyond the first sharp diffraction peak alone, Fig. 4b also shows that the agreement in the structure factor between the 4,096-atom GAP system and experimental data at larger $Q$ is excellent, and significantly better than for the VBSB 100,000-atom system (9).

## Conclusions

Machine-learning-based interatomic potentials can lead to an unprecedented level of quality in the modeling of amorphous materials. In this work, we used a Gaussian approximation potential (GAP) to generate high-quality atomistic structural models of amorphous silicon (*a*-Si), quenching from the liquid at a rate of $10^{11}$ K/s, which has been hitherto inaccessible to DFT-quality simulations. This enabled us to generate structural models of *a*-Si that show convincing agreement with calorimetry and $^{29}$Si NMR experiments, and with X-ray structure factors, including the height of the first sharp diffraction peak. These findings are expected to have implications for future research on disordered and amorphous materials, opening the door for quantitatively accurate atomistic modeling with direct links to experiments, for *a*-Si and beyond.

## Methods

Gaussian approximation potential (GAP) (25) driven simulations were performed using a GAP model for elemental silicon as introduced in recent work (28). This interatomic potential has been fitted to reference DFT data for crystalline, amorphous, and liquid Si configurations, and extensively validated. Among its remarkable features is a correct reproduction of the Si(111)-(7×7) surface reconstruction, which is not possible with state-of-the-art empirical potentials (28).

Simulations were performed using LAMMPS (http://lammps.sandia.gov) (50) and QUIP (https://github.com/libAtoms/QUIP), both of which are interfaced to the GAP framework and freely available for non-commercial research. The time step was 1 fs in all MD simulations. Quenching from the melt was done in the NPT ensemble with a Nosé–Hoover thermostat and a barostat controlling hydrostatic strain (51–53), as implemented in LAMMPS. Simulations started from liquid Si at 1,800 K, followed by equilibration at 1,500 K for 100 ps, and subsequent quenching to 500 K at a constant rate between $10^{14}$ and $10^{11}$ K/s for the 512-atom structures (Figs. 1–3). Further GAP-driven conjugate-gradient (CG) minimizations were performed to relax all atomic positions and cell vectors into local minima, leading to only small changes especially for the $10^{11}$ K/s quench (Supporting Information). Quenching to temperatures lower than 500 K before CG relaxation did not lead to noticeable structural changes. For the 4,096-atom structure (Fig. 4), we quenched with a rate of $10^{13}$ K/s from 1,500 K to 1,250 K, then with $10^{11}$ K/s to 1,050 K, and finally with $10^{13}$ K/s to 500 K, followed by CG relaxation. This is because extensive testing showed that the decisive structural changes occur in the temperature range from 1,250 to 1,050 K, and only there is the (computationally expensive) slow quench rate truly needed.

NMR parameters were computed from first principles, using the DFT-based gauge including projector augmented wave (GIPAW) method (40–42) as implemented in CASTEP (54). We performed self-consistent-field computations with fixed band occupations, using on-the-fly generated ultra-soft pseudopotentials, a 250 eV cutoff energy for the plane-wave expansion, and the PW91 exchange–correlation functional (55). Formally, this method is well-defined only for insulating systems. While finite densities of states at the Fermi level are present in our quenched structures (Supporting Information), these are strongly localized on the isolated threefold-coordinated defect sites; the NMR parameter prediction is therefore made with the assumption of locality. The excellent agreement with experiment for the *a*-Si structure quenched at $10^{11}$ K/s (Fig. 2C and Table 1) supports the validity of this approximation.

To ensure careful relaxation of the other structural models characterized in Fig. 3, these were optimized using GAP in a multistep procedure, including: (i) NVT MD quenches at various volume increments, (ii) a fit of the resulting data to the equation of state, and (iii) a final structural optimization at the so-determined optimum value. The "Tsf+DFT" structure had already been carefully optimized in Ref. (23), and is therefore taken from that study without changes. All structural models generated in this work are provided as Supporting Information.

## Acknowledgements

We thank Professor Jonathan Yates for useful discussions. V.L.D. acknowledges a Feodor Lynen fellowship from the Alexander von Humboldt foundation, a Leverhulme Early Career Fellowship, and support from the Isaac Newton Trust. N.B. acknowledges support from the Office of Naval Research through the U. S. Naval Research Laboratory's basic research base program. A.B.P. acknowledges support from the Collaborative Computational Project for NMR Crystallography, funded by EPSRC grant EP/M022501/1. M.J.C. acknowledges support from Sidney Sussex College, University of Cambridge. L.E.M. acknowledges financial support through a FP7 Marie Curie International Incoming Fellowship. Via our membership of the UK's HEC Materials Chemistry

Consortium, which is funded by EPSRC (EP/L000202), this work used the ARCHER UK National Supercomputing Service (http://www.archer.ac.uk).# References

1. Carlson DE, Wronski CR (1976) Amorphous silicon solar cell. *Appl Phys Lett* 28(11):671–673.

2. Powell MJ (1989) The physics of amorphous-silicon thin-film transistors. *IEEE Trans Electron Devices* 36(12):2753–2763.

3. Street RA (2009) Thin-film transistors. *Adv Mater* 21(20):2007–2022.

4. Cui L-F, Ruffo R, Chan CK, Peng H, Cui Y (2008) Crystalline-amorphous core− shell silicon nanowires for high capacity and high current battery electrodes. *Nano Lett* 9(1):491–495.

5. Key B, et al. (2009) Real-Time NMR Investigations of Structural Changes in Silicon Electrodes for Lithium-Ion Batteries. *J Am Chem Soc* 131(26):9239–9249.

6. Zachariasen W (1932) The Atomic Arrangement in Glass. *J Am Chem Soc* 54(10):3841–3851.

7. Wooten F, Winer K, Weaire D (1985) Computer Generation of Structural Models of Amorphous Si and Ge. *Phys Rev Lett* 54(13):1392–1395.

8. Barkema G, Mousseau N (2000) High-quality continuous random networks. *Phys Rev B* 62(8):4985–4990.

9. Vink RLC, Barkema GT, Stijnman MA, Bisseling RH (2001) Device-size atomistic models of amorphous silicon. *Phys Rev B* 64(24):245214.

10. Fortner J, Lannin JS (1989) Radial distribution functions of amorphous silicon. *Phys Rev B* 39(8):5527–5530.

11. Roorda S, Doorn S, Sinke WC, Scholte PMLO, van Loenen E (1989) Calorimetric Evidence for Structural Relaxation in Amorphous Silicon. *Phys Rev Lett* 62(16):1880–1883.

12. Roorda S, et al. (1991) Structural relaxation and defect annihilation in pure amorphous silicon. *Phys Rev B* 44(8):3702–3725.

13. Laaziri K, et al. (1999) High Resolution Radial Distribution Function of Pure Amorphous Silicon. *Phys Rev Lett* 82(17):3460–3463.

14. Laaziri K, et al. (1999) High-energy x-ray diffraction study of pure amorphous silicon. *Phys Rev B* 60(19):520–533.

15. Griffin JM, et al. (2015) In situ NMR and electrochemical quartz crystal microbalance techniques reveal the structure of the electrical double layer in supercapacitors. *Nat Mater* 14(8):812–819.17

50. Plimpton S (1995) Fast Parallel Algorithms for Short-Range Molecular Dynamics. *J Comput Phys* 117(1):1–19.

51. Parrinello M, Rahman A (1981) Polymorphic transitions in single crystals: A new molecular dynamics method. *J Appl Phys* 52(12):7182–7190.

52. Martyna GJ, Tobias DJ, Klein ML (1994) Constant pressure molecular dynamics algorithms. *J Chem Phys* 101(5):4177–4189.

53. Shinoda W, Shiga M, Mikami M (2004) Rapid estimation of elastic constants by molecular dynamics simulation under constant stress. *Phys Rev B* 69(13):134103.

54. Clark SJ, et al. (2005) First principles methods using CASTEP. *Z Krist* 220:567–570.

55. Perdew JP, Wang Y (1992) Accurate and simple analytic representation of the electron-gas correlation energy. *Phys Rev B* 45(23):13244–13249.